\documentclass[letterpaper, 10 pt, conference]{ieeeconf}

\IEEEoverridecommandlockouts

\usepackage{graphicx}
\usepackage{xcolor}
\usepackage{amsmath}
\usepackage{mathtools}
\usepackage{url}
\usepackage{multirow}
\usepackage{booktabs}
\usepackage{subcaption}
\usepackage{pifont}

\usepackage{hyperref}

\newcommand{\f}[1]{\ensuremath\mathbf{f}_{#1}}

\newcommand{\trademark}{\textsuperscript{TM}}
\newcommand{\best}{\ensuremath\color{red}}

\title{\LARGE \bf
  Bite-Weight Estimation Using Commercial Ear Buds
}

\author{Vasileios Papapanagiotou and Stefanos Ganotakis and Anastasios
  Delopoulos%
  \thanks{Authors are with the Multimedia Understanding Group, Dpt. of
    Electrical and Computer Engineering, Faculty of Engineering, Aristotle
    University of Thessaloniki, Greece {\tt\small vassilis@mug.ee.auth.gr},
    {\tt\small sganotak@ece.auth.gr}, {\tt\small adelo@eng.auth.gr}}%
  \thanks{© 2021 IEEE.  Personal use of this material is permitted. Permission
    from IEEE must be obtained for all other uses, in any current or future
    media, including reprinting/republishing this material for advertising or
    promotional purposes, creating new collective works, for resale or
    redistribution to servers or lists, or reuse of any copyrighted component of
    this work in other works.}%
}

\begin{document}

\maketitle

\begin{abstract}
  While automatic tracking and measuring of our physical activity is a well
  established domain, not only in research but also in commercial products and
  every-day life-style, automatic measurement of eating behavior is
  significantly more limited. Despite the abundance of methods and algorithms
  that are available in bibliography, commercial solutions are mostly limited to
  digital logging applications for smart-phones. One factor that limits the
  adoption of such solutions is that they usually require specialized hardware
  or sensors. Based on this, we evaluate the potential for estimating the weight
  of consumed food (per bite) based only on the audio signal that is captured by
  commercial ear buds (Samsung Galaxy Buds). Specifically, we examine a
  combination of features (both audio and non-audio features) and trainable
  estimators (linear regression, support vector regression, and neural-network
  based estimators) and evaluate on an in-house dataset of 8 participants and 4
  food types. Results indicate good potential for this approach: our best
  results yield mean absolute error of less than 1 g for 3 out of 4 food
  types when training food-specific models, and 2.1 g when training on all
  food types together, both of which improve over an existing literature
  approach.
\end{abstract}

\section{Introduction}
\label{sec:introduction}

In the context of dietary monitoring, various wearable sensors have been
proposed in order to measure different parameters of eating behavior. One of the
first sensors that was used is the in-ear microphone: the in-ear placement
enables the capturing chewing sensors clearly as they are transmitted through
the skull during mastication \cite{amft2005}.

Alternative sensors have also been studied in literature. A piezoelectric sensor
has been used in \cite{sazonov2012}; the sensor is attached on the skin close to
the jaw that captures muscle movement during mastication. The periodic nature of
chewing is also present in the piezoelectric sensor's signal and is used to
detect chewing. Alternative placements of the piezoelectric sensor have also
been examined, such as attached to smart glasses or to neck collars
\cite{kalantarian2014} \cite{farooq2017}. Surface electromyography (EMG) has
also been used for chewing detection \cite{amft2009b} \cite{sonoda2018},
however, is currently one of the least discrete solutions.

Sensors for estimating the weight of a meal (or bite) have also been proposed,
and achieve relatively high effectiveness and low errors. However, the sensors
require manual placement and activation (e.g. plate weight scale
\cite{papapanagiotou2017}) or are part of the table \cite{mattfeld2017} and thus
cannot be used in free-living conditions such as eating outside or on-the-go.

More recently, the interest has been shifting to off-the-shelf solutions to
eliminate the need for specialized hardware as well as decrease the sensors'
intrusiveness. In particular, the 3D accelerometers and gyroscopes that are
commonly embedded in commercial smart-watches can be used to ambiently detect
eating gestures (i.e. the repeated movements of bringing food to the mouth from
a plate, tray, etc) and achieve very promising results in challenging,
free-living conditions \cite{amft2005} \cite{kyritsis2019} \cite{kyritsis2020}
\cite{kyritsis2021}. Alternatively, an accelerometer mounted on the temporalis
\cite{shuangquan2015care,shuangquan2018eating} has also been used to detect
muscle contraction during mastication with promising effectiveness.

Additionally, analysis of photos taken with smart phones can provide detailed
information about eating habits, including types of consumed food, ingredients,
etc (for example, the goFOOD\trademark{} \cite{lu2020gofoodtm} system can
estimate the calorie and macro-nutrient content of a meal based on either two
photos of the meal or a short video). A single photo is used in
\cite{lu2018multi} to perform segmentation, recognition, and volume estimation
of different foods, and results show similar effectiveness to methods that
require multiple photos of the meal.

In this work, we propose a method for estimating bite weight using the audio
signal of commercially available ear buds. Our approach includes extracting
features that are used to train bite weight estimators, based on annotations of
start and stop time-stamps of chews and food type. We evaluate different feature
sets and different types of estimators on an in-house dataset we have collected,
using leave-one-subject-out (LOSO) training and testing. We examine two cases,
one where food type information is available (corresponding to a use-case where
food information is obtained by asking the user directly or by some food-type
recognition system such as \cite{lu2020gofoodtm}) and one where it is not
(corresponding to a use-case where only chewing activity is detected using some
audio-based automated method such as \cite{papapanagiotou2017chewing}).

\section{Related work}
\label{sec:relatedwork}

An algorithm for bite-weight estimation, also from sound captured by an in-ear
microphone, has been proposed in \cite{amft2009}. Audio was recorded at $44$
kHz; a total of $8.64$ hours were recorded from eight individuals.

The algorithm proposed in \cite{amft2009} used $8$ features (Table I of
\cite{amft2009}) that can be extracted from a sequence of chews. Of these $8$
features, $7$ can be computed solely from the start and stop time-stamps of the
chews, and only $1$ requires audio signal (i.e. mean signal energy). For each
chewing bite, these $8$ features are extracted $6$ times: from the entire
chewing bout, from the 1st, 2nd, and 3rd third of the chewing bout, and from the
chewing bouts that consist of the first $3$ and $5$ chews only,
respectively. This yields $48$ features; two more features are computed and a
final vector of $50$ features is produced.

A linear regression (with bias) model is used to estimate bite weight.
A different model is trained for each food type (a total of three food types are
used: potato chip, lettuce, and apple). Two methods of feature selection are
examined: manual selection based on Spearman's correlation coefficient (between
each feature and the bite-weight) and step-wise regression fit. Authors conclude
that both methods yield similar results.

In this work, we differentiate significantly from \cite{amft2009} by (a) using
commercially available ear buds, (b) focusing on audio-based features and
exploring different aggregation methods, and (c) comparing different regression
models for bite weight estimation.

\section{Bite-weight estimation algorithm}
\label{sec:algorithm}

Our proposed approach aims to estimate the weight of a single bite. In short, we
first extract a set of features (we examine both non-audio and audio features)
which we then use to train an estimation model. Models are trained in the
typical LOSO scheme, where a different model is trained for each subject of the
dataset using the data from the other subjects each time.

\subsection{Feature extraction}

\begin{figure}
  \centering
  \includegraphics[scale=.8]{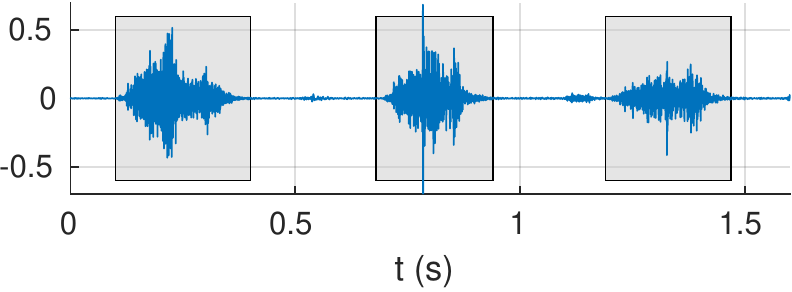}
  \caption{Example of audio signal: the first three chews of an apple bite are
    down, along with the manual ground-truth (gray boxes).}
  \label{fig:chew_example}
\end{figure}

We use two distinct sets of features. The first set (non-audio features) does
not depend on the audio signal, but only on the start and stop time-stamps of
the chews (Figure \ref{fig:chew_example}). Note that in this work the start and
stop time-stamps have been determined manually. Specifically, let $t_{1}[i]$ and
$t_{2}[i]$ for $i=1,\ldots,n$ denote the start and stop time-stamps of a bout of
$n$ chews. We compute the following six features: number of chews (i.e. $n$),
mean and standard deviation of chew duration (chew duration is
$t_{2}[i] - t_{1}[i]$), and mean and standard deviation of chewing rate
(instantaneous chewing rate is estimated as $t_{1}[i] - t_{1}[i - 1]$), and food
type (as is a categorical variable).

The second set of features is based on the audio signal. A challenge lays in the
fact that (a) each chewing bout has a different duration and different number of
chews, (b) each individual chew has a different duration. To overcome this, we
follow a two-step process: first, one feature vector is extracted from each
individual chew of a single chewing bout, and then, all the features vectors of
the chews that belong to a single chewing about are aggregated together to
produce a final, single feature vector for the chewing bout (this final vector
is then used for training the weight predictors).

In the first step, the features that we extract from each individual chew are
the ones used in \cite{papapanagiotou2017,papapanagiotou2020}. The features
include signal energy in log-scale energy bands, higher order statistics
(including skewness and kurtosis), and fractal dimension. Estimating each of
those features is independent of the length of the available audio signal
(i.e. from chew duration); this allows us to obtain comparable values for each
feature among chews of varying duration. After extracting the features for the
entire dataset we standardize them by subtracting the mean (of each feature) and
dividing by its standard deviation.

In the second step, we aggregate the features vectors of the chews of each
chewing bout. We examine two similar approaches to this: bag-of-words (BoW) and
vectors of locally aggregated descriptors (VLAD). Centroids are obtained over
the available training portion of the dataset (AIC is used for selecting the
number of centroids), which are then used both on the train as well as the test
portions of the dataset.

\subsection{Bite-weight estimators}
\label{sec:estimators}

To estimate the bite weight from the available features, we examine four
different algorithms. The first estimator we evaluate is LR, similar to
\cite{amft2009}. We have also experimented with models that include
cross-product terms but have found that the overall effectiveness is not
affected significantly.

The second algorithm is support vector regression (SVR). We use a radial-basis
function (RBF) kernel, and use a grid search for hyper-parameters $C$ and
$\gamma$ by randomly splitting the data from the $m-1$ subjects to $70\%$ for
training and $30\%$ for validation. We search for $C$ in
$10^{i},\,i=-2,-1, 0, 1, 2$ and for $\gamma=10^{i}n^{-1},\,i=-1,0,1,2,3$, where
$n$ is the length of the feature vector.

We also examine classic feed-forward neural networks (FFNN). We consider
architectures with either $2$ or $3$ hidden layers, and $5$, $10$, $15$, and
$20$ neurons per layer (thus, a total of $8$ distinct architectures, as we do
not consider architectures with different number of neurons per layer). The
choice of the architecture is treated as the hyper-parameter of this model and
is selected based on a $90\%$ training and $10\%$ validation split of the $m-1$
subjects, and is thus different for each subject. Training minimizes the mean
absolute error (MAE); learning rate is set to $0.01$ and the maximum number of
epochs is $1,000$. We use the BFGS Quasi-Newton back-propagation algorithm.

Finally, we also examine generalized regression neural networks (GRNN)
\cite{grnn} with Gaussian kernel. Similarly to the previous models, we also
select the hyper-parameter $\sigma$ of the kernel using a train-validation split on
the $m-1$ subjects.

\section{Dataset}
\label{sec:dataset}

To evaluate our approach we have collected an in-house dataset. A total of 8
participants were enrolled for the data collection trials ($6$ males and $2$
females, age $25 \pm 1.07$ years, body-mass index $25.49 \pm 3.06$). Four
different food types were consumed: apple, banana, rice, and potato chips. These
four types were selected as they have a unique combination of crispiness (apple
and potato chips) and wetness (apple and banana). The dataset includes $2$ hours
of eating and contains a total of $473$ chewing bouts and $7.539$ chews.

Audio signals were collected by using commercially available Samsung Galaxy
Buds. We have created a custom Android application that captures synchronized
audio (at $44.1$ kHz, $16$ bit) from the ear buds and plate weight (at $1$ Hz)
from a Bluetooth-enabled plate scale. The plate weight scale has been used to
derive ground truth values for bite weight. A video recording of each session
has also been captured to further assist us in the annotation process.

\section{Evaluation}
\label{sec:evaluation}

To evaluate our approach we perform various combinations of feature sets and
estimation models. Given the non-audio based feature set and the two methods of
aggregating the audio based features, we examine the following five
combinations: non-audio features ($\f{1}$), audio with BoW ($\f{2}$), audio with
VLAD ($\f{3}$), combination of non-audio and audio with BoW features ($\f{4}$),
and combination of non-audio and audio with VLAD features ($\f{5}$). For each of
these feature sets we examine four estimators (Section \ref{sec:estimators}):
LR, SVR, FFNN, GRNN. Finally, we train five different models per combination:
the first four are food specific, while the fifth is trained on the data from
all food types. Table \ref{tab:result_mae} shows the mean absolute error per
experiment, and Table \ref{tab:result_mape} shows the mean absolute percentage
error (\%) in the same structure. All experiments are performed in LOSO fashion;
thus, each result is the mean across the $8$ participants of our dataset.

\begin{table}
	\centering
	\caption{Mean and standard deviation absolute errors per algorithm and
      feature set. Best result (lowest error) in red.}
	\label{tab:result_mae}
	\begin{tabular}{lccccc}
      \toprule
      & \textbf{Apple} & \textbf{Banana} & \textbf{Rice} & \textbf{Chips} & \textbf{All} \\
      \midrule
      & \multicolumn{5}{l}{\textit{LR}} \\
      $\f{1}$ & $3.02\,(2.3)$ & $5.89\,(2.8)$ & $3.80\,(2.5)$ & $0.93\,(0.7)$ & $4.94\,(3.4)$ \\
      $\f{2}$ & $4.13\,(3.4)$ & $5.82\,(3.9)$ & $3.82\,(2.2)$ & $1.23\,(0.8)$ & $3.70\,(3.2)$ \\
      $\f{3}$ & $4.83\,(3,7)$ & $5.57\,(3.4)$ & $5.06\,(3.2)$ & $1.10\,(0.8)$ & $3.86\,(3.4)$ \\
      $\f{4}$ & $3.65\,(2.5)$ & $5.54\,(3.3)$ & $3.86\,(3.1)$ & $1.02\,(0.7)$ & $3.24\,(2.8)$ \\
      $\f{5}$ & $3.15\,(2.2)$ & $5.97\,(3.3)$ & $3.74\,(2.7)$ & $0.92\,(0.7)$ & $3.44\,(3.0)$ \\
      \midrule
      & \multicolumn{5}{l}{\textit{SVR}} \\
      $\f{1}$ & $3.18\,(2.2)$ & $6.30\,(3.2)$ & $3.97\,(2.7)$ & $0.97\,(0.8)$ & $4.52\,(3.3)$ \\
      $\f{2}$ & $4.76\,(3.5)$ & $5.01\,(3.2)$ & $4.44\,(3.0)$ & $1.17\,(0.8)$ & $3.66\,(3.2)$ \\
      $\f{3}$ & $5.07\,(3.3)$ & $5.11\,(3.2)$ & $5.02\,(3.2)$ & $1.12\,(0.8)$ & $3.76\,(3.5)$ \\
      $\f{4}$ & $3.57\,(2.3)$ & $5.56\,(3.2)$ & $3.19\,(2.6)$ & $1.05\,(0.8)$ & $3.16\,(2.9)$ \\
      $\f{5}$ & $3.00\,(2.4)$ & $5.44\,(3.4)$ & $3.68\,(2.7)$ & $1.00\,(0.8)$ & $3.02\,(2.5)$ \\
      \midrule
      & \multicolumn{5}{l}{\textit{FFNN}} \\
      $\f{1}$ & $3.70\,(2.7)$ & $5.17\,(3.6)$ & $4.72\,(3.2)$ & $0.93\,(0.8)$ & $3.31\,(3.0)$ \\
      $\f{2}$ & $4.19\,(3.7)$ & $4.68\,(3.8)$ & $4.29\,(3.4)$ & $1.00\,(0.9)$ & $3.86\,(3.4)$ \\
      $\f{3}$ & $3.64\,(2.6)$ & $4.10\,(3.2)$ & $3.52\,(2.1)$ & $0.81\,(0.6)$ & $4.10\,(3.4)$ \\
      $\f{4}$ & $3.56\,(2.7)$ & $5.56\,(3.5)$ & $4.52\,(3.0)$ & $1.22\,(0.9)$ & $2.77\,(2.7)$ \\
      $\f{5}$ & $2.60\,(2.4)$ & $2.55\,(3.0)$ & $2.22\,(2.4)$ & $\best{} 0.20\,(0.4)$ & $\best{} 2.12\,(2.4)$ \\
      \midrule
      & \multicolumn{5}{l}{\textit{GRNN}} \\
      $\f{1}$ & $2.39\,(1.9)$ & $3.82\,(2.3)$ & $3.27\,(2.1)$ & $0.82\,(0.6)$ & $3.59\,(2.8)$ \\
      $\f{2}$ & $3.17\,(3.0)$ & $2.85\,(2.2)$ & $2.50\,(2.4)$ & $0.87\,(0.8)$ & $4.14\,(3.6)$ \\
      $\f{3}$ & $4.43\,(4.0)$ & $4.78\,(2.9)$ & $4.01\,(2.8)$ & $1.04\,(0.7)$ & $6.19\,(3.4)$ \\
      $\f{4}$ & $\best{} 1.10\,(1.8)$ & $\best{} 0.89\,(1.1)$ & $\best{} 0.93\,(1.4)$ & $0.25\,(0.4)$ & $3.90\,(3.4)$ \\
      $\f{5}$ & $4.30\,(3.2)$ & $5.66\,(3.8)$ & $3.61\,(2.9)$ & $1.08\,(0.8)$ & $3.80\,(3.4)$ \\
      \midrule
      & \multicolumn{5}{l}{\textit{Amft et al.}} \\
      & $3.26\,(2.7)$ & $5.96\,(2.9)$ & $5.22\,(3.2)$ & $1.10\,(0.8)$ & $3.37\,(3.0)$ \\
      \bottomrule
	\end{tabular}
\end{table}

\begin{table}
	\centering
	\caption{Mean and standard deviation of absolute relative (\%) errors per
      algorithm and feature set. Best result (lowest error) in red.}
	\label{tab:result_mape}
	\begin{tabular}{lccccc}
      \toprule
      & \textbf{Apple} & \textbf{Banana} & \textbf{Rice} & \textbf{Chips} & \textbf{All} \\
      \midrule
      & \multicolumn{5}{l}{\textit{LR}} \\
      $\f{1}$ & $31\,(27)$ & $54\,(36)$ & $47\,(46)$ & $47\,(59)$ & $57\,(26)$ \\
      $\f{2}$ & $40\,(39)$ & $57\,(43)$ & $44\,(45)$ & $60\,(73)$ & $38\,(34)$ \\
      $\f{3}$ & $46\,(44)$ & $44\,(41)$ & $51\,(53)$ & $49\,(34)$ & $40\,(47)$ \\
      $\f{4}$ & $38\,(29)$ & $50\,(37)$ & $47\,(51)$ & $40\,(49)$ & $32\,(36)$ \\
      $\f{5}$ & $28\,(27)$ & $55\,(40)$ & $43\,(42)$ & $37\,(44)$ & $35\,(32)$ \\
      \midrule
      & \multicolumn{5}{l}{\textit{SVR}} \\
      $\f{1}$ & $28\,(24)$ & $56\,(38)$ & $48\,(48)$ & $47\,(47)$ & $53\,(59)$ \\
      $\f{2}$ & $47\,(40)$ & $42\,(38)$ & $43\,(40)$ & $54\,(43)$ & $38\,(33)$ \\
      $\f{3}$ & $44\,(36)$ & $42\,(38)$ & $47\,(48)$ & $50\,(39)$ & $40\,(43)$ \\
      $\f{4}$ & $37\,(27)$ & $52\,(39)$ & $38\,(46)$ & $40\,(49)$ & $31\,(32)$ \\
      $\f{5}$ & $27\,(25)$ & $51\,(38)$ & $42\,(41)$ & $48\,(39)$ & $31\,(36)$ \\
      \midrule
      & \multicolumn{5}{l}{\textit{FFNN}} \\
      $\f{1}$ & $33\,(25)$ & $46\,(42)$ & $50\,(38)$ & $40\,(28)$ & $33\,(29)$ \\
      $\f{2}$ & $36\,(60)$ & $40\,(42)$ & $41\,(41)$ & $50\,(45)$ & $40\,(29)$ \\
      $\f{3}$ & $33\,(58)$ & $29\,(21)$ & $36\,(45)$ & $42\,(42)$ & $46\,(41)$ \\
      $\f{4}$ & $30\,(26)$ & $56\,(43)$ & $45\,(44)$ & $56\,(58)$ & $39\,(31)$ \\
      $\f{5}$ & $26\,(26)$ & $26\,(32)$ & $25\,(35)$ & $\best{} 15\,(28)$ & $\best{} 20\,(27)$ \\
      \midrule
      & \multicolumn{5}{l}{\textit{GRNN}} \\
      $\f{1}$ & $22\,(18)$ & $36\,(30)$ & $40\,(38)$ & $34\,(52)$ & $35\,(31)$ \\
      $\f{2}$ & $28\,(25)$ & $19\,(17)$ & $28\,(38)$ & $35\,(57)$ & $42\,(37)$ \\
      $\f{3}$ & $36\,(38)$ & $32\,(32)$ & $44\,(58)$ & $51\,(54)$ & $64\,(37)$ \\
      $\f{4}$ & $\best{} 09\,(14)$ & $\best{} 06\,(10)$ & $\best{} 09\,(20)$ & $20\,(45)$ & $40\,(43)$ \\
      $\f{5}$ & $43\,(31)$ & $57\,(50)$ & $42\,(47)$ & $44\,(46)$ & $36\,(37)$ \\
      \midrule
      & \multicolumn{5}{l}{\textit{Amft et al.}} \\
      & $30\,(23)$ & $54\,(33)$ & $60\,(53)$ & $55\,(73)$ & $34\,(31)$ \\
      \bottomrule
	\end{tabular}
\end{table}

We also present evaluation results from our implementation of the algorithm of
\cite{amft2009} (Section \ref{sec:relatedwork}).

\begin{figure}
  \centering
  \includegraphics[scale=.8]{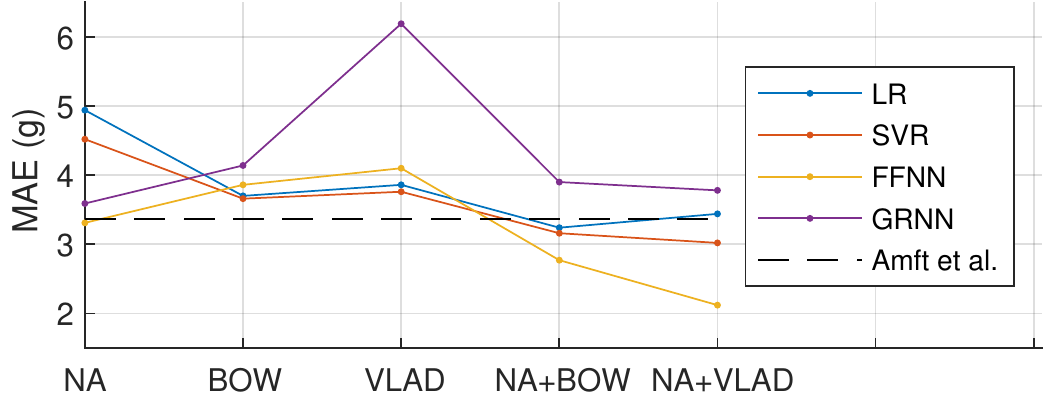}
  \caption{Mean absolute error per feature set and estimator model when training
    on all food types together.}
  \label{fig:results1}
\end{figure}

Based on the results, the combination of both non-audio and audio based features
improves the estimation accuracy (yielding lower error) for most cases. This is
more evident when training on all food types together (Figure
\ref{fig:results1}). This conclusion is also inline with the results of the
algorithm of \cite{amft2009} which uses a combination of $7$ non-audio features
and $1$ audio feature, as it is better from our non-audio and audio-only
approaches while slightly worse from our non-audio and audio combinations.

Based on the results, FFNN and GRNN are able to achieve the best results (lowest
errors) compared to LR and SVR. When training on a single food type, GRNN-based
models with $\f{4}$ achieve the lowest MAE (close to or less than $1$ g) and
similarly low standard deviation of absolute errors. The only exception is for
potato chips where FFNN achieves the lowest errors. However, GRNN achieves the
second lowest errors and the difference (from FFNN)) is very small:
$0.25\,(0.4)$ g for GRNN versus $0.20\,(0.4)$ for FFNN. When training on all
food types together, FFNN with $\f{5}$ seems to achieve the lowest errors.

Comparing the two different types of aggregating audio features (from chews to
chewing bouts) there seems to be no clear conclusion about whether BoW is better
of VLAD. This holds both for when using only audio features (i.e. $\f{2}$
vs. $\f{3}$), as well as for when combining them with the non-audio features
(i.e. $\f{4}$ vs. $\f{5}$). The only exception is GRNN that seems to benefit
from the use of BoW; this can be attributed to the stricter quantization of the
feature space that BoW applies.

Finally, MAE is quite lower for potato chips compared to the other food
types. However, this is not a result of ``better'' trained estimation models,
but of the fact that potato-chip bites are generally lighter (than apple bites
for example). This can be confirmed by comparing MAPE errors that are shown in
Table \ref{tab:result_mape}.


Evaluation results for the algorithm of Amft et al. \cite{amft2009} are
comparable with our approach. The are clearly surpassed though by our FFNN and
GRNN based approaches. Overall, MAE is higher in our dataset compared to the
values reported by the authors in their original work of \cite{amft2009}. This
can be attributed to the more challenging nature of our dataset as well as
differences in the sound captured by our commercially available ear buds and
their custom-made sensor.

\section{Conclusions}
\label{sec:conclusions}

In this work we have presented an approach for estimating bite weight from audio
signal captured by commercially available ear buds. Using commercially available
hardware is essential to enable higher adoption rates for such dietary
monitoring approaches, since they reduce invasiveness and discomfort of the end
user.

Our approach uses a combination of non-audio and audio features which are used
to train estimation models. We evaluate on an in-house dataset of approximately
$2$ hours. Our best results are obtained by training food-specific GRNN models
and non-food-specific FFNN models. GRNN models yield MAE of approximately $1$ g
or less, and FFNN yield a total MAE of $2.12$ g. We also compare with an
existing algorithm from literature and achieve lower errors for all cases.

An important limitation of our approach is that it requires annotations for the
start and stop time-stamps of individual chews, as well as food type annotations
(for food-type--specific models). Future work includes evaluating on bigger and
more diverse datasets with more food types and different data-capturing
conditions (closer to free-living) as well as evaluating in combination with
audio based chewing detectors and automatically detected food types.

\section*{Acknowledgments}

The work leading to these results has received funding from the EU Commission
under Grant Agreement No. 965231, the REBECCA project (H2020).

\bibliographystyle{IEEEtran}
\bibliography{IEEEabrv, root}

\end{document}